\documentclass[usenatbib]{mn2e}

\usepackage{aas_macros}
\usepackage{amsmath}
\usepackage{amssymb}
\usepackage[colorlinks=true,breaklinks=true,citecolor=blue]{hyperref} 
\usepackage{xspace}
\usepackage{color}
\usepackage{graphicx}
\usepackage{txfonts} 
\usepackage{mathtools} 
\usepackage{tensor}
\usepackage{multirow}

\newcommand{\eg}{\mbox{e.\,g.,}\xspace} 
\newcommand{\Eg}{\mbox{E.\,g.,}\xspace}
\newcommand{\ie}{\mbox{i.\,e.,}\xspace}
\newcommand{\Ie}{\mbox{I.\,e.,}\xspace}

\renewcommand{\vec}[1]{\bmath{#1}} 

\newcommand{\angstrom}{\textup{\AA}}

\definecolor{blue}{rgb}{0,0,1} 
\definecolor{yellow}{rgb}{1,0.84,0}
\definecolor{red}{rgb}{1,0,0}
\definecolor{mypink1}{rgb}{0.858, 0.188, 0.478}
\definecolor{darkgreen}{rgb}{0,0.5,0}
\definecolor{orange}{rgb}{1.0,0.4,0}

\title[High column density absorbers and the Lyman-alpha forest]{Simulating the effect of high column density absorbers on \\the one-dimensional Lyman-alpha forest flux power spectrum}
\author[K.~K.~Rogers et al.]
{Keir K.~Rogers,$^{1}$\thanks{E-mail: keir.rogers.14@ucl.ac.uk} Simeon Bird,$^{2,5}$ Hiranya V.~Peiris,$^{1,3}$ Andrew Pontzen,$^{1}$
\newauthor Andreu Font-Ribera$^{1}$ and Boris Leistedt$^{4,5}$
\\
$^1$Department of Physics \& Astronomy, University College London, Gower Street, London WC1E 6BT, UK\\
$^2$Department of Physics \& Astronomy, Johns Hopkins University, 3400 N.~Charles Street, Baltimore, MD 21218, USA\\
$^3$Oskar Klein Centre for Cosmoparticle Physics, Stockholm University, AlbaNova, Stockholm SE-106 91, Sweden\\
$^4$Department of Physics, New York University, 726 Broadway, New York, NY 10003, USA\\
$^5$NASA Einstein Fellow}

\begin{document}
\maketitle
\begin{abstract}
We measure the effect of high column density absorbing systems of neutral hydrogen ({\sc Hi}) on the one-dimensional (1D) Lyman-alpha forest flux power spectrum using cosmological hydrodynamical simulations from the Illustris project. High column density absorbers (which we define to be those with {\sc Hi} column densities \(N(\textsc{Hi}) > 1.6 \times 10^{17}\,\mathrm{atoms}\,\mathrm{cm}^{-2}\)) cause broadened absorption lines with characteristic damping wings. These damping wings bias the 1D Lyman-alpha forest flux power spectrum by causing absorption in quasar spectra away from the location of the absorber itself. We investigate the effect of high column density absorbers on the Lyman-alpha forest using hydrodynamical simulations for the first time. We provide templates as a function of column density and redshift, allowing the flexibility to accurately model residual contamination, \ie if an analysis selectively clips out the largest damping wings. This flexibility will improve cosmological parameter estimation, \eg allowing more accurate measurement of the shape of the power spectrum, with implications for cosmological models containing massive neutrinos or a running of the spectral index. We provide fitting functions to reproduce these results so that they can be incorporated straightforwardly into a data analysis pipeline.
\end{abstract}

\section{Introduction}
\label{sec:intro}

The Lyman-alpha forest (a series of neutral hydrogen absorption lines in the spectra of quasars) is a uniquely powerful probe of the clustering of matter at redshifts from about $z = 2$ to  $z = 6$ \citep{1998ApJ...495...44C,2000ApJ...543....1M,2005ApJ...635..761M,2013PhRvD..88d3502V,2013A&A...559A..85P,2016MNRAS.tmp.1599I} and from sub-Mpc to hundreds of Mpc scales. The one-dimensional (1D) Lyman-alpha forest flux power spectrum (along the line of sight) is particularly sensitive to small-scale clustering in the quasi-linear regime and provides important constraints on  extended cosmological models that suppress small-scale power \citep{2005PhRvD..71j3515S,2015JCAP...11..011P,2017arXiv170201764I,2017arXiv170203314Y,2017arXiv170304683I,2017arXiv170309126A}, notably those containing massive neutrinos and warm dark matter. This small-scale information complements the larger scales probed by the angular power spectrum of the cosmic microwave background (CMB). For example, the best upper limit on the sum of neutrino masses \citep{2015JCAP...11..011P} comes from combining CMB data from the Planck Collaboration \citep{2016A&A...594A..11P} with the 1D Lyman-alpha forest power spectrum as measured from Sloan Digital Sky Survey (SDSS)-III/Baryon Oscillation Spectroscopic Survey (BOSS) Data Release 9 (DR9) quasar spectra \citep{2011AJ....142...72E,2013AJ....145...10D,2013A&A...559A..85P}.

Future surveys like the Dark Energy Spectroscopic Instrument \citep[DESI;][]{2016arXiv161100036D,2016arXiv161100037D} will further improve constraints on extended cosmological models. \citet{2014JCAP...05..023F} forecast one-sigma errors on a DESI measurement of the sum of neutrino masses to be 0.017 eV\footnote{This is the full forecasted constraint considering a combination of \textit{Planck} CMB data, DESI broadband galaxy power spectrum, DESI broadband Lyman-alpha forest flux power spectrum and \(\sim 100\) high-resolution Lyman-alpha forest quasar spectra.}. Considering that the lower limit on the sum of neutrino masses from neutrino oscillation experiments is 0.06 eV \citep{2014JHEP...11..052G,2014PhRvD..90i3006F,2017JHEP...01..087E}, this would constitute at least a three-sigma detection. Furthermore, the 1D Lyman-alpha forest flux power spectrum probes the primordial power spectrum on the smallest currently accessible scales, $k \sim 4$ Mpc$^{-1}$. Including Lyman-alpha forest data will improve constraints on the running of the spectral index (which quantifies deviations from a pure power-law spectrum) by a factor of two, reaching one-sigma errors of $\pm 0.002$ \citep{2014JCAP...05..023F}. This would provide new insights into early universe physics, potentially ruling out classes of models of inflation. Importantly, it will also provide a unique independent cross-check at small scales of the primordial power spectrum shape inferred from CMB measurements at large scales.

Achieving these limits requires marginalisation over the uncertain impact of a number of astrophysical effects on the 1D Lyman-alpha forest power spectrum. In particular, this includes broadened absorption features from high column density absorbers. High column density absorbers are usually classified as either damped Lyman-alpha absorbers (DLAs), with column densities \(N(\textsc{Hi})\) exceeding \(2 \times 10^{20}\,\mathrm{atoms}\,\mathrm{cm}^{-2}\) \citep{1986ApJS...61..249W}, or Lyman-limit systems (LLS), which correspond to \(2 \times 10^{20}\,\mathrm{atoms}\,\mathrm{cm}^{-2} > N(\textsc{Hi}) > 1.6 \times 10^{17}\,\mathrm{atoms}\,\mathrm{cm}^{-2}\). Both types of system produce broad damping wings which extend large distances in redshift space. If not accounted for, they will bias cosmological parameter estimation from the Lyman-alpha forest. The systems are formed at peaks of the underlying density distribution; consequently, they cluster more strongly than the forest itself \citep{2012JCAP...11..059F}.

To remove the bias induced by damped absorbers, one can fit a model for their effect on power spectra. The most widely used approach \citep{2005MNRAS.360.1471M} is now more than a decade old. Although this model was adequate for the data available at the time, future surveys will be substantially more constraining and therefore demand tighter control over systematics. Furthermore, there have been significant improvements in theoretical modelling of these systems \citep[\eg][]{pontzen08,2015MNRAS.447.1834B}. An updated model for the effects of high column density absorbers is therefore both timely and essential in order to achieve the forecasted cosmological constraints from future surveys.

Different column densities correspond to gas at different physical densities, so that simulations suitable for modelling the forest are often not suited to reproducing high column density systems. The Lyman-alpha forest is largely insensitive to the physics of galaxy formation since it is sourced by gas at below mean density; the primary uncertainties arise from cosmological parameters and the thermal history of the intergalactic medium. Conversely, high column density absorbers arise largely from regions within or around galaxies and are thus very sensitive to the physics of galaxy formation and less sensitive to large-scale cosmology. It is consequently essential to model the effect of high column density absorbers using simulations which include detailed galaxy formation physics and can thus reproduce their characteristics and statistics.

In Lyman-alpha forest studies, damping wings are sometimes ``clipped'' (\ie removed or masked) from quasar spectra \citep[\eg see][for details of the process for BOSS DR9 spectra]{2013AJ....145...69L}. However, not all damping wings are identified and many will remain in the spectra, especially in noisier spectra where they are harder to spot and for lower-density absorbers (\ie LLS) which have narrower wings. Therefore, in the final cosmological parameter estimation from the 1D Lyman-alpha forest power spectrum, the effect of residual high column density absorbers is modelled as a multiplicative scale-dependent bias of the power spectrum with an amplitude (reflecting the level of residual contamination) that is fitted and marginalised \citep{2015JCAP...11..011P}. The functional form of this model (\ie its scale and redshift dependence) is based on the measurements made in \citet{2005MNRAS.360.1471M}.

\citet{2005MNRAS.360.1471M} investigated the effect with lognormal model mock quasar spectra \citep[\ie generated without hydrodynamical simulations; details of their generation are given in][]{2006ApJS..163...80M}, since the numerical simulations available at the time were not large enough to generate spectra encompassing the full width of damping wings. They then probe the effect of high column density absorbers on the Lyman-alpha forest by inserting damping wings in mock spectra at the peaks of the lognormal field, based on the observationally-determined column density distribution function (CDDF). They find a systematic effect on the observed 1D Lyman-alpha forest power spectrum that is maximised on scales corresponding to the width of a damped system and which has negligible redshift evolution (considering three redshift slices at \(z = [2.2,3.2,4.2]\)). They provide a single template to fit their bias measurement, including the effect of all LLS and DLAs together. However, as discussed above, in current data analysis pipelines, damping wings are removed from quasar spectra in a way that preferentially removes higher density systems. Therefore, when the template is used in parameter inference, it may not correctly model the bias of the \textit{residual} contamination, which will have a different CDDF to the total --- the clipping of the survey spectra changes the survey CDDF. The bias will have a different scale-dependence (not just amplitude), since this is driven by the distribution of the widths of damping wings remaining in quasar spectra.

In this work, we investigate the effect of high column density absorbers on the 1D Lyman-alpha forest power spectrum as a function of their column density and redshift using hydrodynamical simulations of galaxy formation from the Illustris project \citep{2014Natur.509..177V,2015A&C....13...12N}. Comparison to relevant observations has shown that Illustris reproduces the observed CDDF and spatial clustering of high-density systems \citep[][see \S~\ref{sec:sims} for more details]{2014Natur.509..177V,2014MNRAS.445.2313B} at the 95\% confidence level. Spectra are generated from this simulation, then separated into categories according to the maximum column density within each spectrum (see \S~\ref{sec:hcd_absorbers} for more details). We measure the 1D flux power spectrum of each of these types of spectrum and measure the (multiplicative) bias of each type compared to the power spectrum of the Lyman-alpha forest alone. We make this measurement at multiple redshifts and so probe the redshift evolution of this effect.

We discuss high column density absorbers in more detail in \S~\ref{sec:hcd_absorbers}. In \S~\ref{sec:method}, our methodology in going from hydrodynamical simulations to measurements of the 1D flux power spectrum is explained. We present our main results in \S~\ref{sec:results}. These results are discussed in \S~\ref{sec:discussion} and in \S~\ref{sec:templates}, we present the templates that we have fitted to our measurements. Finally, conclusions are drawn in \S~\ref{sec:concs}.

\section{Damped Lyman-alpha absorbers and Lyman-limit systems}
\label{sec:hcd_absorbers}

\begin{table*}\centering
\caption{The neutral hydrogen ({\sc Hi}) column density limits \([N(\textsc{Hi})_\mathrm{min}, N(\textsc{Hi})_\mathrm{max}]\) that define the categories of absorbing systems used in this work. The columns on the right show the percentage of spectra (at each redshift \(z\) that is considered) in our \((106.5\,\mathrm{Mpc})^3\) simulation box \citep[][Illustris-1]{2014Natur.509..177V,2015A&C....13...12N} where the highest-density system belongs to a given category.}
\label{tab:hcd_absorbers}
\begin{tabular}{cccccccc}
\hline
\multirow{2}{*}{Absorber category} & \(N(\textsc{Hi})_\mathrm{min}\) & \(N(\textsc{Hi})_\mathrm{max}\) & \multicolumn{5}{c}{\% of spectra in \((106.5\,\mathrm{Mpc})^3\) simulation at} \\
 & \multicolumn{2}{c}{[\(\mathrm{atoms}\,\mathrm{cm}^{-2}\)]} & \(z = 2.00\) & \(z = 2.44\) & \(z = 3.01\) & \(z = 3.49\) & \(z = 4.43\) \\
\hline
Lyman-alpha forest & 0 & \(1.6 \times 10^{17}\) & 77.7 & 69.6 & 57.4 & 45.7 & 22.0 \\
LLS & \(1.6 \times 10^{17}\) & \(1 \times 10^{19}\) & 10.6 & 14.9 & 21.8 & 27.0 & 36.6 \\
Sub-DLA & \(1 \times 10^{19}\) & \(2 \times 10^{20}\) & 5.9 & 8.1 & 11.4 & 14.3 & 20.1 \\
Small DLA & \(2 \times 10^{20}\) & \(1 \times 10^{21}\) & 3.1 & 4.1 & 5.5 & 7.8 & 12.8 \\
Large DLA & \(1 \times 10^{21}\) & \(\infty\) & 2.7 & 3.3 & 3.9 & 5.2 & 8.5 \\
\hline
\end{tabular}
\end{table*}

High column density absorbers are regions of neutral hydrogen ({\sc Hi}) gas that are above a column density threshold of \(N(\textsc{Hi}) > 1.6 \times 10^{17}\,\mathrm{atoms}\,\mathrm{cm}^{-2}\). By contrast, lower column density absorbers form the Lyman-alpha forest. The absorption lines formed by high column density absorbers are broadened, forming damping wings and hence absorption in the spectrum away from the location of the absorbing gas. The damping wings have a characteristic Voigt profile, which is a convolution of a Gaussian profile (caused by Doppler broadening) and a Lorentzian profile (caused by natural or collision broadening). The width of these wings in velocity space increases with the column density of the absorbing system. High column density absorbers are then usually classified as either damped Lyman-alpha absorbers (DLAs), whose damping wings are considered significantly broadened and which correspond to \(N(\textsc{Hi}) > 2 \times 10^{20}\,\mathrm{atoms}\,\mathrm{cm}^{-2}\) \citep{1986ApJS...61..249W}; or Lyman-limit systems (LLS), which correspond to column densities in the range \(2 \times 10^{20}\,\mathrm{atoms}\,\mathrm{cm}^{-2} > N(\textsc{Hi}) > 1.6 \times 10^{17}\,\mathrm{atoms}\,\mathrm{cm}^{-2}\).

In this work, we aim to investigate the effect of high column density absorbers (and especially their damping wings) on the one-dimensional Lyman-alpha forest flux power spectrum, as a function of their column density (and redshift). We therefore use a more refined classification of high column density absorbers based on their column densities, in particular accounting for the fact that higher density LLS do have wide damping wings. Table \ref{tab:hcd_absorbers} shows the column density limits that define our categories, as well as the percentage of simulated spectra (see \S~\ref{sec:sims}) where the highest-density system is a given type and hence is the main contaminant. The overall percentage of spectra contaminated by high column density absorbers (LLS, sub-DLAs, small and large DLAs) increases with redshift because the {\sc Hi} CDDF increases at higher densities at higher redshifts, but always there are more LLS than DLAs.

\section{Method}
\label{sec:method}

We first outline the method we have used and then explain the steps in more detail in the following subsections (\S~\ref{sec:sims} to \ref{sec:hcd_bias}).
\begin{enumerate}
\renewcommand{\theenumi}{(\arabic{enumi})}
\item We use a cosmological hydrodynamical simulation from the Illustris project \citep{2014Natur.509..177V,2015A&C....13...12N} and generate mock spectra on a grid (562\,500 in total, each at a velocity resolution of \(25\,\mathrm{km}\,\mathrm{s}^{-1}\) and with a typical length of $\simeq 8\,000\,\mathrm{km}\,\mathrm{s}^{-1}$). We repeat this for a number of redshift slices at which the Lyman-alpha forest is observed (\(z = [2.00,2.44,3.01,3.49,4.43]\)). (See \S~\ref{sec:sims}.)
\item For each redshift slice, we separate the spectra according to the highest column density system within that spectrum using the absorber categories defined in Table \ref{tab:hcd_absorbers}. For each absorber category (and the total set of spectra), we measure the one-dimensional (1D) flux power spectrum (\ie along the line of sight, integrating over transverse directions) using a fast Fourier transform (FFT). (See \S~\ref{sec:pow_spectrum}.)
\item We then measure the (multiplicative) bias of the flux power spectra from each category relative to the 1D flux power spectrum of the Lyman-alpha forest, as a function of absorber type (\ie maximum column density) and redshift (see \S~\ref{sec:hcd_bias}). We fit parametric models to these bias measurements and provide these templates in \S~\ref{sec:templates}. 
\end{enumerate}

\subsection{Hydrodynamical simulations and mock spectra}
\label{sec:sims}

Our main results make use of snapshots from the highest-resolution (in terms of both  dark matter particles and hydrodynamical cells) cosmological hydrodynamical simulation from the Illustris project \citep[][Illustris-1\footnote{The simulation we use is publically available at \url{http://www.illustris-project.org/data}.}]{2014Natur.509..177V,2015A&C....13...12N}. The simulation adopts the following cosmological parameters: \(\Omega_\mathrm{m} = 0.2726\), \(\Omega_\Lambda = 0.7274\), \(\Omega_\mathrm{b} = 0.0456\), \(\sigma_8 = 0.809\), \(n_\mathrm{s} = 0.963\) and \(H_0 = 100\,h\,\mathrm{km}\,\mathrm{s}^{-1}\,\mathrm{Mpc}^{-1}\), where \(h = 0.704\) \citep{2014MNRAS.444.1518V}. The box has a volume in comoving units of \((106.5\,\mathrm{Mpc})^3\) and we consider snapshots at redshifts \(z = [2.00,2.44,3.01,3.49,4.43]\).

The Illustris simulations use the moving mesh code \texttt{AREPO} \citep{2010MNRAS.401..791S}. The galaxy formation physics implemented is of relevance to dense regions of neutral hydrogen gas, and therefore we describe it briefly here. The subgrid models include prescriptions for supernova \citep{2003MNRAS.339..289S,2013MNRAS.436.3031V} and active galactic nuclei (AGN) \citep{2005MNRAS.361..776S,2007MNRAS.380..877S} feedback (\citealt{2014MNRAS.445.2313B} showed that the properties of DLAs are quite insensitive to the details of AGN feedback); radiative cooling; star formation and metal enrichment of gas. Self-shielding is implemented as a correction to the photoionization rate, which is a function of hydrogen density and gas temperature. The potential ionising effect of local stellar radiation within the most dense absorbers (\ie large DLAs) \citep[\eg][]{2011MNRAS.418.1796F} is neglected. \citet{2010MNRAS.402.1523P} found this effect to be negligible and accurate calculations in any case require physics on parsec scales, well below the resolution of the simulation (it can then be viewed as part of the unresolved physics included in the above feedback prescriptions). More details of these models are given in \citet{2013MNRAS.436.3031V,2014MNRAS.445.2313B}. Gravitational interactions are computed using the TreePM approach \citep{2005MNRAS.364.1105S}.

We require that these simulations accurately reproduce the necessary statistics of high column density absorbers that are observed in surveys. As a means of quantifying this, we can first consider the CDDF of neutral hydrogen over relevant column densities (\(N(\textsc{Hi}) > 1.6 \times 10^{17}\,\mathrm{atoms}\,\mathrm{cm}^{-2}\)). \citet{2014Natur.509..177V} make a comparison of the CDDF as produced by Illustris centered at \(z = 3\) to the distribution observed in a number of surveys [\citet{2010ApJ...718..392P} for LLS; \citet{2013A&A...556A.141Z} for sub-DLAs; \citet{2009A&A...505.1087N} for DLAs]. In particular, the distributions are consistent with the feature in the CDDF around the DLA threshold, where the distribution rises, being reproduced well (the results of \citealt{2017MNRAS.466.2111B} from SDSS-III DR12 spectra are also consistent for DLAs). \citet{2014MNRAS.445.2313B} showed that the \texttt{AREPO} code with the above hydrodynamical models can produce values of the DLA halo bias (at \(z = 2.3\)) which are in agreement with measured values from real surveys \citep{2012JCAP...11..059F}, indicating that the clustering of high column density absorbers is well reproduced. \citet{2015MNRAS.447.1834B} compared the distribution function of velocity widths of low ionization metal absorbers associated with DLAs as produced by the simulations at \(z = 3\) to the distribution observed in \citet{2013ApJ...769...54N}. The data points are within the 68\% confidence interval of the simulated distribution. This suggests that the simulations are reproducing the kinematics, and thus the host halo distribution, of high column density absorbers. One potential caveat is that these simulations are found to produce too high a total incidence rate of DLAs when compared to observations \citep{2012A&A...547L...1N} at \(z = 2\) \citep{2014MNRAS.445.2313B}. However, the overall incidence rate is absorbed into a normalisation that must in any case be allowed to float during analysis of clipped spectra (as discussed in \S~\ref{sec:discussion}).

For each snapshot, we generate mock spectra containing only the Lyman-alpha absorption line (\ie with a rest wavelength of \(1215.67 \angstrom\)) from neutral hydrogen. We do this on a square grid of 562\,500 spectra, in the plane perpendicular to a direction that we define as the line of sight. Each spectrum extends the full length of the simulation box with periodic boundary conditions, giving a size in velocity (or ``redshift'') space of \(\{7111, 7501, 8000, 8420, 9199\}\,\mathrm{km}\,\mathrm{s}^{-1}\) respectively at \(z = [2.00,2.44,3.01,3.49,4.43]\)\footnote{We convert the comoving length of the box to a proper velocity by the Hubble law.}. We first measure the optical depth \(\tau\) in velocity bins of size \(25\,\mathrm{km}\,\mathrm{s}^{-1}\) along the spectrum\footnote{For comparison, BOSS DR9 spectra are binned at a velocity resolution of \(69.02\,\mathrm{km}\,\mathrm{s}^{-1}\) \citep{2013AJ....145...69L}.}. We further convolve our spectra with a Gaussian kernel of \(\mathrm{FWHM} = 8\,\mathrm{km}\,\mathrm{s}^{-1}\), setting the simulated spectrographic resolution. We then calculate the transmitted flux \(\mathcal{F} = e^{-\tau}\). In this way, the spectra we have constructed are insensitive to contamination from other absorption (or emission) lines, estimation of the emitted quasar continuum (which here is effectively set to unity) or instrumental noise. In each spectrum pixel, we are also able to measure the column density (integrated along the line of sight in each bin) of neutral hydrogen, which we use in measuring the maximum density systems in each spectrum (\S~\ref{sec:pow_spectrum}).

\subsection{One-dimensional flux power spectrum}
\label{sec:pow_spectrum}

We separate our spectra into the absorber categories (Lyman-alpha forest, LLS, sub-DLAs, small and large DLAs) defined in Table \ref{tab:hcd_absorbers} according to the maximum column density system within each spectrum. We search for the highest column density integrated over any four neighbouring velocity bins; this amounts to a comoving length along the line of sight of \(\{1.50, 1.42, 1.33, 1.27, 1.16\}\,\mathrm{Mpc}\) respectively at \(z = [2.00,2.44,3.01,3.49,4.43]\). The categorisation is insensitive to the number of neighbouring velocity bins that we use, as the boundaries between categories differ by orders of magnitude in column density. Moreover, the method is efficient in identifying high column density absorbers since they are vastly more dense than the surrounding gas forming the Lyman-alpha forest\footnote{We have explicitly tested the impact of doubling or halving the number of neighbouring velocity bins we use on the 1D flux power spectra we measure in each absorber category. We find that the maximum absolute difference in any power spectrum bin is a negligible 0.2\%.}. We have chosen a length that is much larger than the most extensive DLAs as found by recent studies \citep{2012MNRAS.424L...1K} and so we are sure to integrate over the full length of any high column density absorbers. Our definition of high column density absorbers includes blends, where a number of smaller, lower column density systems have been added together. In this way, we have associated with each spectrum the most dominant absorbing system and in the case where high column density absorbers are identified, these are the main contamination to the spectrum through their associated damping wings. The percentage of spectra in each absorber category at each redshift slice is given in Table \ref{tab:hcd_absorbers}.

We measure the 1D flux power spectrum of all the spectra and each absorber category at each redshift slice. The 1D power spectrum \(P^\mathrm{1D}(k_{||}, z)\) is defined as the integral of the three-dimensional (3D) power spectrum \(P^\mathrm{3D}(k_{||}, \vec{k_\perp}, z)\) over directions perpendicular to the line of sight:
\begin{equation}\label{eq:power_1D}
P^\mathrm{1D}(k_{||}, z) = \int P^\mathrm{3D}(k_{||}, \vec{k_\perp}, z)\, \frac{\mathrm{d}\vec{k_\perp}}{(2\pi)^2},
\end{equation}
where the wavevector \(\vec{k} = [k_{||}, \vec{k_\perp}]\) is conjugate to velocities in real space and so is measured in units of inverse velocity (\eg \(\mathrm{s}\,\mathrm{km}^{-1}\)). We also use the convention of absorbing the \(2 \pi\) into the conjugate variable\footnote{\Ie we define the Fourier transform as \(\delta(k) = \int \delta(x) e^{-ikx} \mathrm{d}x\).}.

To measure $P^{\mathrm{1D}}$ for an individual line of sight, we first calculate the fluctuation in each velocity \(v_{||}\) bin \(\delta_\mathcal{F}(v_{||}) = \frac{\mathcal{F}(v_{||})}{\langle \mathcal{F} \rangle} - 1\), where \(\langle \mathcal{F} \rangle\) is the average flux over all spectra at each redshift \citep{1998ApJ...495...44C}. We calculate the 1D Fourier transform along the line of sight \(\hat{\delta}_\mathcal{F} (k_{||})\) using a fast Fourier transform (FFT)-based method since we have evenly-spaced velocity bins. We then estimate the 1D flux power spectrum for each sightline \(P_\mathrm{Raw}^\mathrm{1D}(k_{||}) = |\hat{\delta}_\mathcal{F} (k_{||})|^2\). Finally, we estimate the 1D flux power spectrum in Eq.~\eqref{eq:power_1D} for each absorber category \(i\) by \citep[\eg][]{2013A&A...559A..85P}
\begin{equation}\label{eq:power_FT}
P_i^\mathrm{1D}(k_{||}, z) = \left\langle \frac{P_\mathrm{Raw}^\mathrm{1D}(k_{||}, z)}{W^2(k_{||}, \Delta v, R)} \right\rangle_i,
\end{equation}
where we explicitly indicate that the raw 1D power spectra depend on redshift \(z\). The average is taken over spectra of a given category (or all spectra for the total power spectrum) at each redshift slice. The window function \(W(k_{||}, \Delta v, R)\) that is divided out arises from the binning in velocity space (\(\Delta v\)) and the simulated spectrographic resolution \(R\):
\begin{equation}\label{eq:window_func}
W(k_{||}, \Delta v, R) = \exp\,\left(- \frac{1}{2} (k_{||} R)^2\right) \times \frac{\sin\,(k_{||} \Delta v / 2)}{k_{||} \Delta v / 2},
\end{equation}
where \(\Delta v = 25\,\mathrm{km}\,\mathrm{s}^{-1}\) and \(R = 3.40\,\mathrm{km}\,\mathrm{s}^{-1}\) (not to be confused with the spectrographic resolving power; see \S~\ref{sec:sims}). We then have an estimate of the 1D flux power spectrum for each absorber category of spectra at each redshift slice.

\subsection{Modelling the effect of high column density absorbers}
\label{sec:hcd_bias}

The total 1D flux power spectrum of a set of spectra \(P_\mathrm{Total}^\mathrm{1D} (k_{||}, z)\) can be expressed as a weighted sum of the 1D flux power spectra calculated in Eq.~\eqref{eq:power_FT} for each absorber category \(i\):
\begin{equation}\label{eq:weighted_sum}
P_\mathrm{Total}^\mathrm{1D} (k_{||}, z) = \sum_i \alpha_i (z)\,P_i^\mathrm{1D} (k_{||}, z),
\end{equation}
where \(\alpha_i (z)\) are the fraction of spectra in each absorber category at each redshift (as given in Table \ref{tab:hcd_absorbers} for our simulated ensemble of spectra). In a real survey, \(\alpha_i (z)\) may change from their raw values due to the attempt to clip (\ie remove) high column density absorbers discussed in \S~\ref{sec:intro}. We can rearrange Eq.~\eqref{eq:weighted_sum} to isolate the 1D flux power spectrum of the Lyman-alpha forest alone:
\begin{equation}\label{eq:multiplicative_bias}
P_\mathrm{Total}^\mathrm{1D} (k_{||}, z) = P_\mathrm{Forest}^\mathrm{1D} (k_{||}, z)\,\left[\alpha_\mathrm{Forest} (z) + \sum_{i \neq \mathrm{Forest}} \alpha_i (z) \frac{P_i^\mathrm{1D} (k_{||}, z)}{P_\mathrm{Forest}^\mathrm{1D} (k_{||}, z)}\right].
\end{equation}
In this way, we have isolated the effect of spectra containing high column density absorbers on the 1D flux power spectrum of the Lyman-alpha forest as a multiplicative bias (\ie the terms in square brackets)\footnote{We could simplify this form further by asserting the fact that \(\sum_i \alpha_i (z) = 1\) to remove the parameter \(\alpha_\mathrm{Forest} (z)\), but it is useful to keep this form as we explain in \S~\ref{sec:templates}.}. This matches the general form of modelling this effect in previous studies, as explained in \citet{2015JCAP...11..011P} (based on the results in \citealt{2005MNRAS.360.1471M}), but now additionally probing the bias as a function of column density (\ie by using the different absorber categories). We discuss in more detail in \S~\ref{sec:discussion} our motivations for using this particular form of the bias (as opposed to \eg an additive bias). Using the 1D flux power spectra we have calculated in \S~\ref{sec:pow_spectrum}, we are able to measure the fractions in Eq.~\eqref{eq:multiplicative_bias} (\(P_i^\mathrm{1D} (k_{||}, z) / P_\mathrm{Forest}^\mathrm{1D} (k_{||}, z)\)) and we present the results in \S~\ref{sec:results}.

\section{Results}
\label{sec:results}

\begin{figure}
\includegraphics[width=\columnwidth]{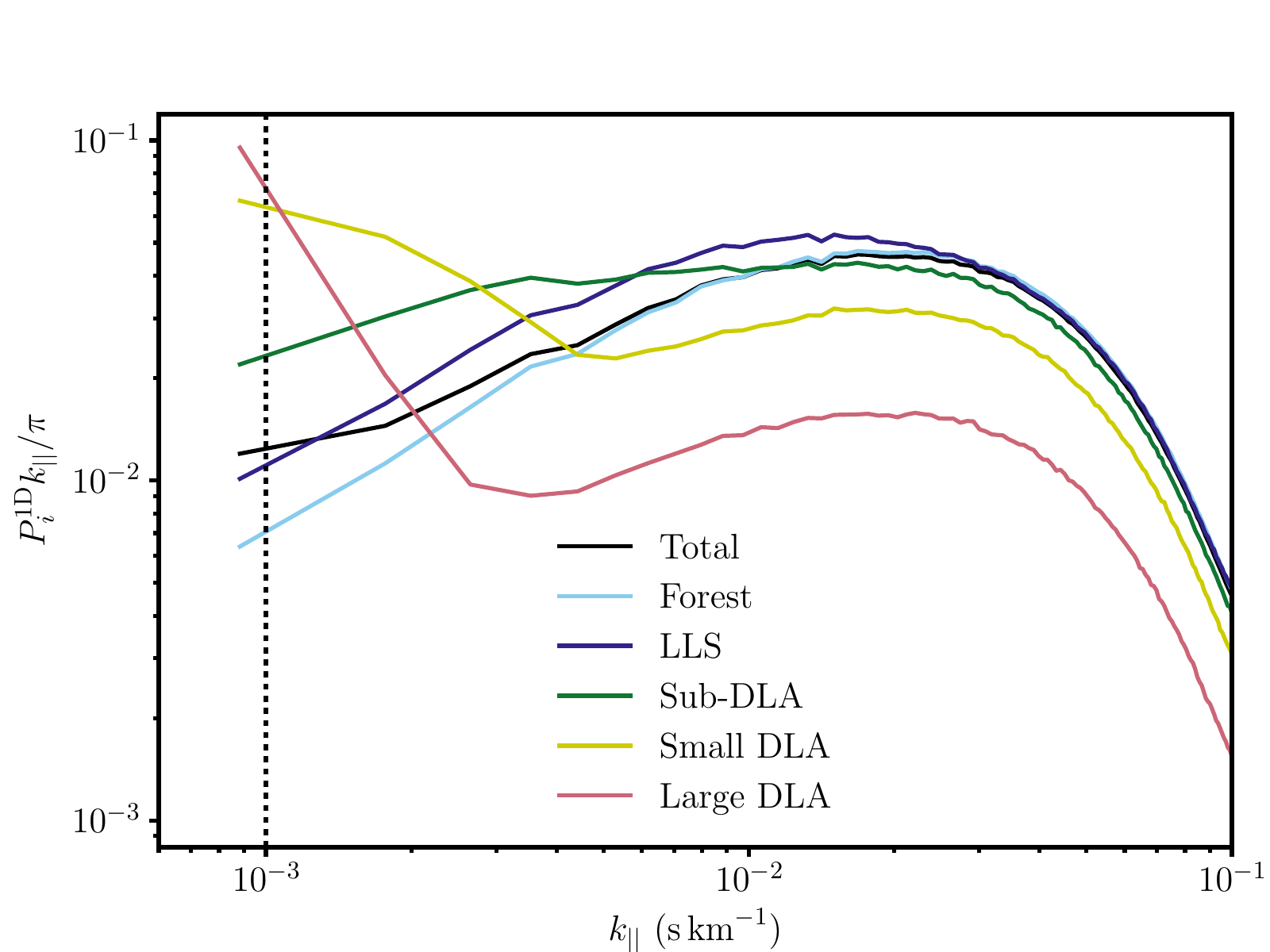}
\caption{The one-dimensional flux power spectra of different categories of spectra, as a function of line-of-sight scale \(k_{||}\) at redshift \(z = 2.00\). The different categories are: the total from our full simulated sample of spectra; spectra containing only Lyman-alpha forest; and spectra contaminated by different types of high column density absorber [LLS, sub-DLAs, small and large DLAs]. The vertical dashed line shows the largest scale probed by the BOSS DR9 1D Lyman-alpha forest flux power spectrum; by comparison, the largest scale probed by our analysis at this redshift is larger at \(9 \times 10^{-4}\,\mathrm{s}\,\mathrm{km}^{-1}\). The definitions of the different categories of  absorber are given in Table \ref{tab:hcd_absorbers}. (See \S~\ref{sec:templates} for the full intermediate redshift evolution.)}
\label{fig:contaminant_power}
\end{figure}

Figure \ref{fig:contaminant_power} shows the 1D flux power spectra of different subsets of sightlines that we have measured from our simulations [see \S~\ref{sec:pow_spectrum} and in particular Eq.~\eqref{eq:power_FT}] at redshift \(z = 2.00\). The different subsets shown are: the total as would be measured if no distinction between different types of spectra was made; spectra containing only Lyman-alpha forest (\ie the ensemble that is uncontaminated by high column density absorbers); and spectra contaminated by different categories of high column density absorber, as defined in Table \ref{tab:hcd_absorbers}. We first note that the total 1D flux power spectrum at any redshift slice can be reconstructed as a weighted sum of the other 1D flux power spectra for each absorber category at that redshift (see \S~\ref{sec:hcd_bias} and in particular Eq.~\eqref{eq:weighted_sum}). The weights are the fraction of spectra in each category (the values we measure for our simulated ensemble are given in Table \ref{tab:hcd_absorbers}). We can estimate the fractional (\(1 \sigma\)) statistical error on each power spectrum data-point as \(1 / \sqrt{N_i}\), where \(N_i\) is the number of input modes (\ie simulated spectra) per data-point \(i\). This assumes that data-points and input modes are independent. This is largest for the large DLA power spectrum at \(z = 2.00\), which has 15,188 input simulated spectra giving an error of 0.81\%, and smallest for the forest power spectrum at \(z = 2.00\), which has 437,063 input simulated spectra, giving an error of 0.15\%. All the other uncertainties for each measured power spectrum range in-between these values and can be computed from Table \ref{tab:hcd_absorbers}.

The total power spectrum deviates from the Lyman-alpha forest power spectrum at all redshifts, showing there is a bias from contamination of spectra by high column density absorbers. This bias can be deconstructed as a function of column density by looking at the power spectra of different absorber categories. The power spectra of high column density absorbers have a distinctive increase on large scales (small \(k_{||}\)). This is caused by self-correlations across the width of damping wings, which (as discussed in \S~\ref{sec:hcd_absorbers}) can be modelled by a Voigt profile (a convolution of a Gaussian and a Lorentzian). Therefore, the power spectrum of high column density absorbers (on large scales) is connected to the Fourier transform of a Voigt profile. This increases for higher column density systems since there is more line broadening, and starts on larger scales for higher column density systems since the damping wings are wider. (See Appendix \ref{sec:voigt} for more analysis and discussion of the Voigt profile model.)  On small scales, all the power spectra converge to a scaled version of the Lyman-alpha forest flux power spectrum. This reflects the fact that contaminated spectra do contain some uncontaminated spectral pixels. The amplitude of the small-scale power spectrum reflects the fraction of spectra that is uncontaminated, increasing for lower-column density systems since their damping wings are narrower. There is some sensitivity to the length of our simulated spectra, which primarily manifests in our results as the amplitude of the small-scale residual Lyman-alpha forest power in the contaminated power spectra. This is because longer simulated contaminated spectra would have a larger fraction with residual Lyman-alpha forest. This is discussed further and explicitly modelled such that this effect is removed in \S~\ref{sec:templates}.

\begin{figure}
\includegraphics[width=\columnwidth]{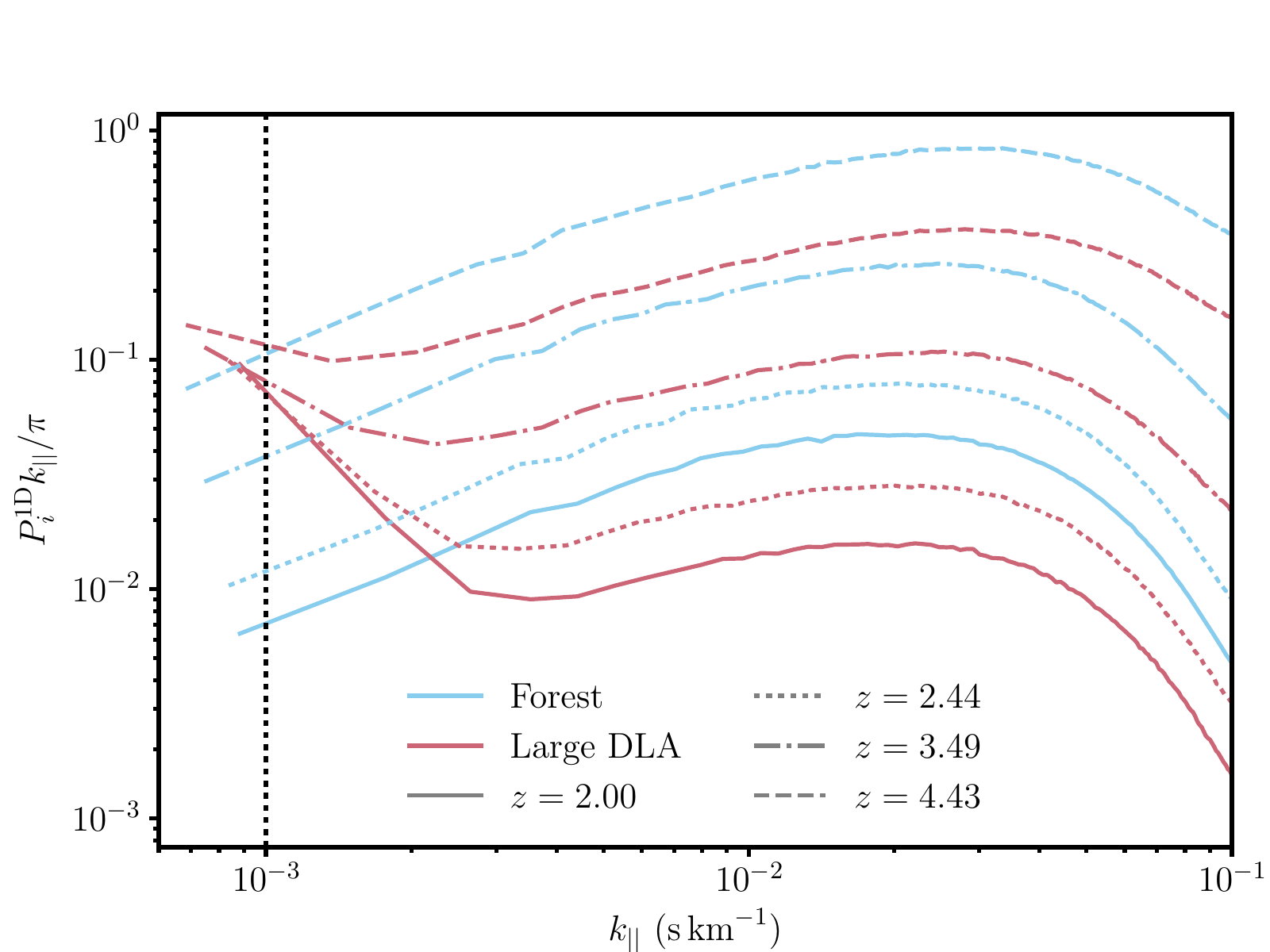}
\caption{As Fig.~\ref{fig:contaminant_power}, but showing more of the redshift slices that we consider (for \(z = [2.00, 2.44, 3.49, 4.43]\)), for spectra containing only Lyman-alpha forest and spectra contaminated by large DLAs.}
\label{fig:contaminant_power_z_evol}
\end{figure}

Figure \ref{fig:contaminant_power_z_evol} shows 1D flux power spectra as in Fig.~\ref{fig:contaminant_power}, but for more of the redshift slices that we consider (\(z = [2.00, 2.44, 3.49, 4.43]\)), for spectra containing only Lyman-alpha forest and spectra contaminated by large DLAs. The Lyman-alpha forest flux power spectrum has the expected shape, amplitude and redshift evolution, matching observations \citep[\eg][]{2013A&A...559A..85P} and reflecting the fact that it is an integral of a (biased) matter power spectrum. A peculiarity of the Lyman-alpha forest flux power spectrum is that its amplitude \textit{increases} with redshift (unlike the linear matter power spectrum); this is because neutral hydrogen is more abundant at higher redshift and so there is more absorption in quasar spectra (\ie the Lyman-alpha forest becomes a more negatively biased tracer of the matter distribution). By contrast, it can be seen that the large-scale correlations associated with the large DLAs are converging to a single point as redshift changes. This reflects the fact that these correlations arise from individual damping wings, which do not evolve with redshift.

\begin{figure}
\includegraphics[width=\columnwidth]{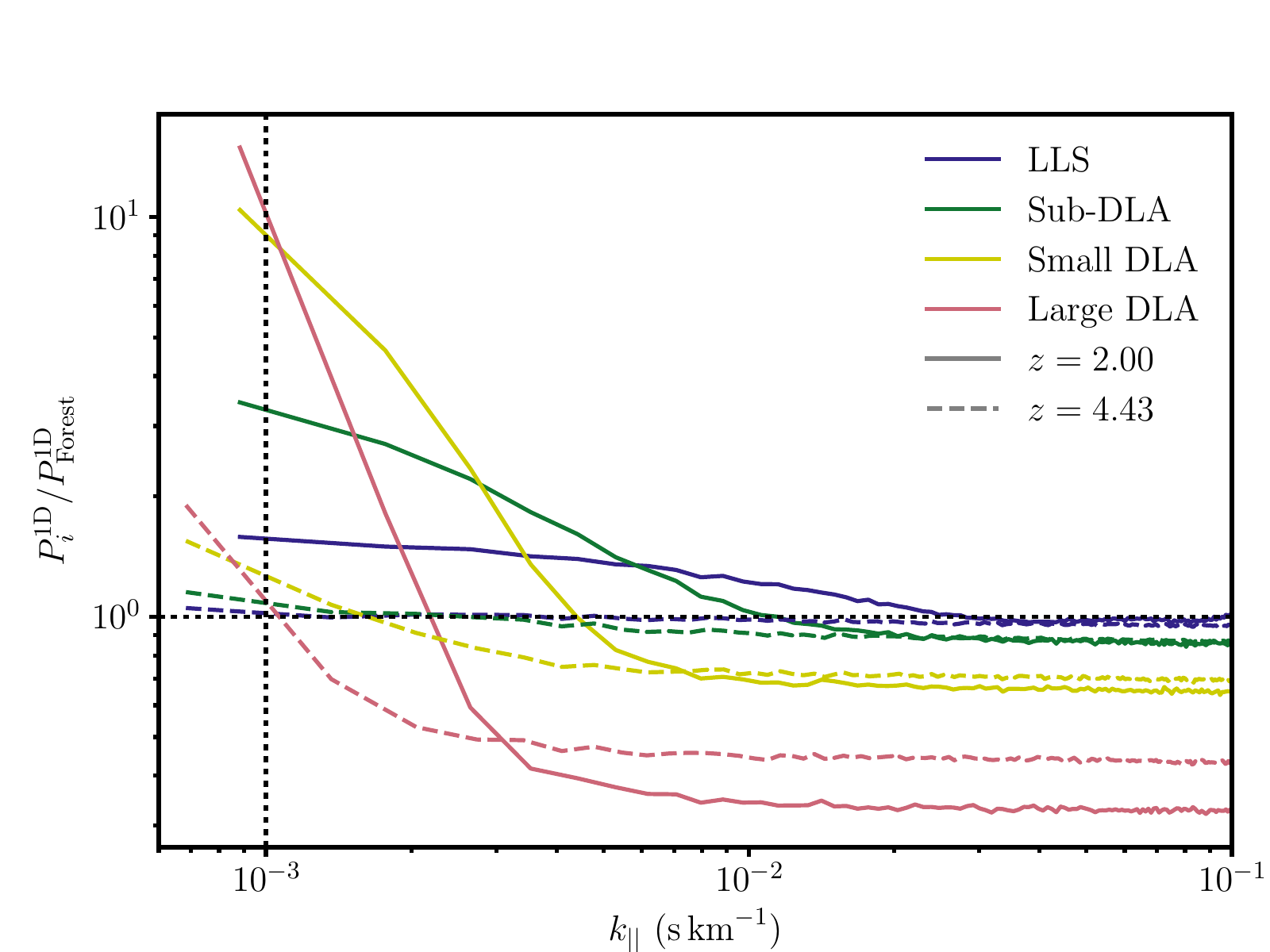}
\caption{The multiplicative bias of high column density absorbers to the one-dimensional Lyman-alpha forest flux power spectrum, as a function of line-of-sight scale \(k_{||}\) and redshift \(z\), \ie the ratio of the 1D flux power spectrum of spectra contaminated by high column density absorbers [LLS, sub-DLAs, small and large DLAs] over spectra containing only Lyman-alpha forest. The vertical dashed line shows the largest scale probed by the BOSS DR9 1D Lyman-alpha forest flux power spectrum. The definitions of the different categories of high column density absorber are given in Table \ref{tab:hcd_absorbers}. The different line styles correspond to different redshift slices, showing the maximum and minimum redshifts that we consider. (See \S~\ref{sec:templates} for the full intermediate redshift evolution.)}
\label{fig:contaminant_ratios}
\end{figure}

Figure \ref{fig:contaminant_ratios} shows the same 1D flux power spectra as in Figs.~\ref{fig:contaminant_power} and \ref{fig:contaminant_power_z_evol}, but now as ratios between the flux power of spectra contaminated by high column density absorbers and the flux power of spectra containing only Lyman-alpha forest, for \(z = 2.00\) and \(z = 4.43\). These ratios are the quantities to which we fit our templates (see \S~\ref{sec:templates}) as part of our bias model (see \S~\ref{sec:hcd_bias}). Plotted in this form, it is clear that the large-scale corrections associated with damping wings increase with column density of the damped system. The corrections also decrease with increasing redshift because the Lyman-alpha forest flux power spectrum (on the denominator of the ratio) increases with redshift. On small scales, the ratios converge to a constant value, which reflects the fraction of a line of sight that is uncontaminated (see above). The redshift evolution of this constant value  is driven by the transformation from comoving to velocity space: spectra are longer in velocity space at higher redshift (despite being drawn from the same comoving length of the simulation). Conversely, the width of damping wings does not change with redshift (for a given column density) because this width just arises from the physical processes within the hydrogen gas (rather than cosmological evolution). Therefore, the fraction of spectra uncontaminated by the damping wings increases with redshift.

\section{Discussion}
\label{sec:discussion}

We first discuss and summarise the results we have presented in \S~\ref{sec:results}. Using our measurements from cosmological hydrodynamical simulations, we have been able to confirm and characterise the effect of high column density absorbers on the 1D Lyman-alpha forest flux power spectrum as a function of column density, scale and redshift. There are distinctive large-scale correlations across the widths of individual damping wings (a ``one-halo'' term) arising from high column density absorbers that are seen to bias the 1D flux power spectrum of a set of spectra, relative to the power spectrum of the Lyman-alpha forest alone (Fig.~\ref{fig:contaminant_power}). These correlations persist for all the high column density absorbing systems that we identify (\ie for all column densities \(N(\textsc{Hi}) > 1.6 \times 10^{17}\,\mathrm{atoms}\,\mathrm{cm}^{-2}\)). Our results can be further understood by relating the shape and amplitude of the large-scale power spectrum of spectra contaminated by high column density absorbers to the Fourier transform of the Voigt profile that is normally used to model damping wings (due to the combination of physical effects that broaden absorption lines; see Appendix \ref{sec:voigt}). We find evidence in our simulation results that the 1D flux power spectrum of high column density absorbers does not evolve with redshift (Fig.~\ref{fig:contaminant_power_z_evol}). This reflects the fact that the Voigt profiles of damping wings depend only on column density (\ie the physical processes within high column density absorbing regions) and not redshift (\ie cosmological evolution) (see Eq.~\eqref{eq:voigt_tau}).

The most recent previous investigation into the effect of high column density absorbers on the Lyman-alpha forest was performed by \citet{2005MNRAS.360.1471M} \citep[see also][]{1999ApJ...520....1C,2004MNRAS.349L..33V}. These authors measured a single bias function for the 1D Lyman-alpha forest flux power spectrum (at each redshift they consider) that includes the combined effect of all high column density absorbers (\ie all LLS and DLAs). Our results are qualitatively similar to those of the previous study; however, by investigating different absorber categories based on column density ranges (Table \ref{tab:hcd_absorbers}), we have shown that the form of the bias as a function of wavenumber depends strongly on column density. 

This will have implications for any parameter inference from the 1D flux power spectrum. For instance, the analysis by \citet{2015JCAP...11..011P} uses a single multiplicative bias model for the Lyman-alpha forest flux power spectrum based on the results in \citet{2005MNRAS.360.1471M}\footnote{The model is reported in \citet{2015JCAP...11..011P} as \(1 - 0.2\,\alpha_\mathrm{DLA}\,[1 / (15000.0\,k_{||} - 8.9) + 0.018]\), where \(\alpha_\mathrm{DLA}\) is the free amplitude.}. The model has a free amplitude that is allowed to vary (reflecting the level of contamination in a given survey) and is then marginalised. The shape of this model is therefore based on the observed CDDF of high column density absorbers. However, as discussed in \S~\ref{sec:intro}, in the measurement of the 1D flux power spectrum, high column density absorbers in the quasar spectra are clipped out in the hope of removing noise \citep{2013AJ....145...69L,2013A&A...559A..85P}. This process changes the CDDF of high column density absorbers by preferentially removing higher column density systems which are easier to spot in the noisy spectra. Hence, the shape of the bias from residual high column density absorbers is different (as we have shown in \S~\ref{sec:results}) and the model used by \citet{2013A&A...559A..85P} may not be flexible enough to account for this, especially at the level of accuracy required by future surveys.  Our measurements provide a set of templates for the effect of different absorber categories as a function of column density. By using our templates as part of the model in Eq.~\eqref{eq:multiplicative_bias}, it is now possible to more accurately characterise the bias of the residual contamination. We also find evidence for redshift dependence of the fractional effect of high column density absorbers on the forest power spectrum (driven by the changing amplitude of the forest power spectrum), which is also not included in the current model, but is incorporated into our templates. Fits allowing incorporation of our new results into future pipelines are described in \S~\ref{sec:templates}.

We now discuss our motivations for some of the choices made in our analysis. We have chosen to present our main results as the 1D flux power spectra of different sets of simulated spectra, where we have categorised spectra according to the maximum column density system within each spectrum. This means that we are measuring the power spectra of ensembles of spectra that are contaminated to similar extents, rather than the flux power spectra of high column density absorbers alone. Furthermore, a consequence of this categorisation is that within the spectra of a given category, there may be high column density absorbers of a lower column density (\eg there may be LLS in the large-DLA category of spectra). In the first instance, this does not affect our results because the power spectrum measurements we have made (\S~\ref{sec:results}) and the templates that we construct (\S~\ref{sec:templates}) include the effect of this possible additional lower column density contamination. A subtlety arises because the amount of additional lower column density contamination will be partly sensitive to the length of simulated spectra, since longer spectra have a greater chance of being contaminated. However, the damping wings of the highest column density systems already produce zero transmitted flux over a significant fraction of the length of our simulation box, so that the presence of possible additional high column density absorbers will make very little difference in any case. We tested this by inserting an LLS into a spectrum contaminated by a large DLA, which reduced the total transmitted flux by 0.07\%. By carrying out this insertion test with a ``control'' scenario without the additional contamination, we are able to show that this subtlety will have negligible impact on our conclusions and the validity of our templates.

Finally, we comment on the particular form of our preferred model for the bias of high column density absorbers to the 1D Lyman-alpha forest flux power spectrum (as shown in Eq.~\ref{eq:multiplicative_bias}). We model the bias as a multiplicative correction, rather than \eg an additive form. First, this matches the form of the currently-used model \citep[as shown in][]{2015JCAP...11..011P}. Moreover, an additive form would require either the separation of high column density absorbers and the Lyman-alpha forest in the simulated spectra or a complete physical understanding of how the two components interact at the ensemble level. The former is not trivial for our analysis since we are not inserting high column density absorbers (as previous studies have done), but are simultaneously simulating the low and high column density regions of gas. We avoid the latter due to any remaining physical uncertainties and instead form a parametric multiplicative model based on our simulated results (see \S~\ref{sec:templates}).

\section{Templates for the effect of high column density absorbers}
\label{sec:templates}

\begin{figure}
\includegraphics[width=\columnwidth]{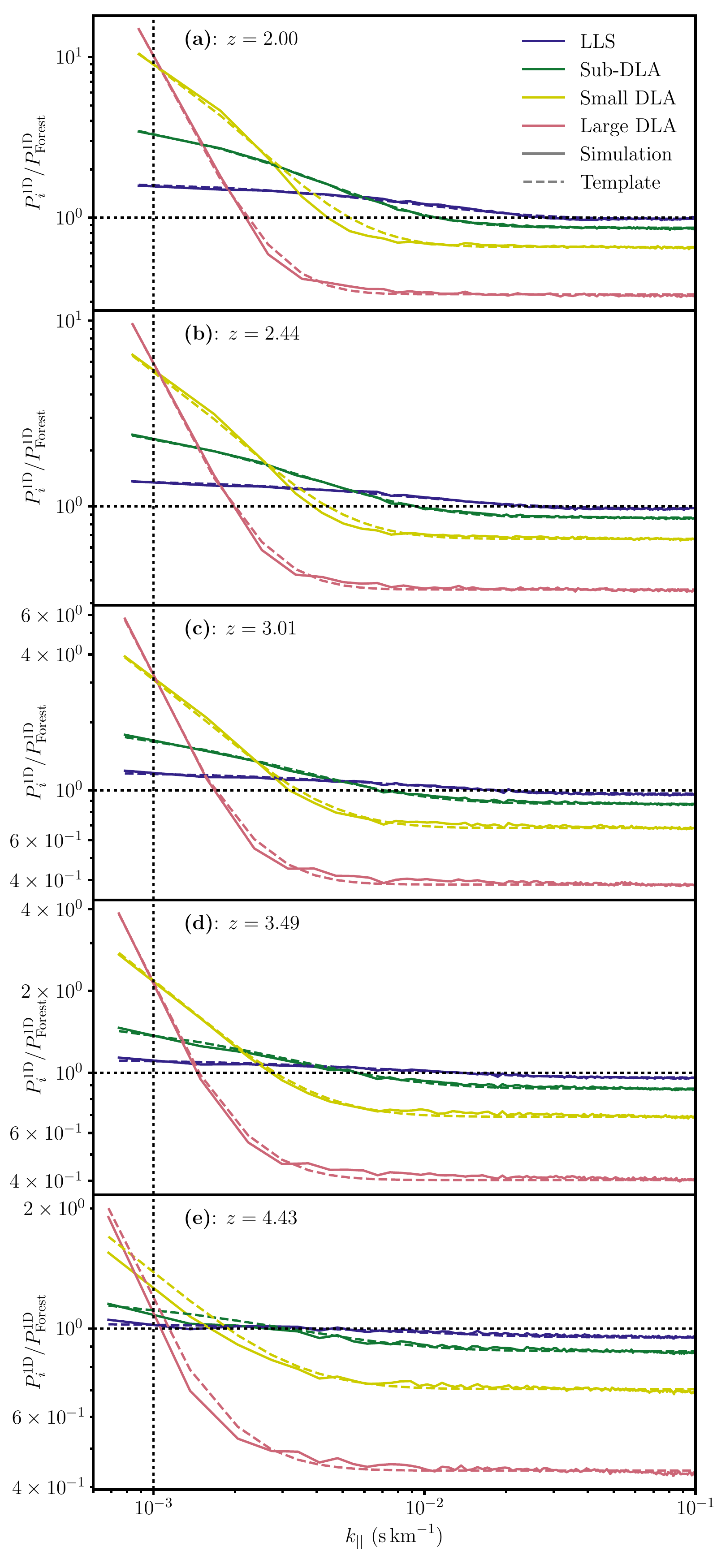}
\caption{The multiplicative bias of high column density absorbers to the one-dimensional Lyman-alpha forest flux power spectrum, as a function of line-of-sight scale \(k_{||}\) and redshift \(z\), \ie the ratio of the 1D flux power spectrum of spectra contaminated by high column density absorbers [LLS, sub-DLAs, small and large DLAs] over spectra containing only Lyman-alpha forest. The solid lines are these ratios as measured in the hydrodynamical simulations; the dashed lines are our best-fitting templates to these measurements. The functional form of our templates is given in Eq.~\eqref{eq:parametric_model} and the best-fit values of the model parameters are given in Table \ref{tab:template_param_vals}. The vertical dashed lines show the largest scale probed by the BOSS DR9 1D Lyman-alpha forest flux power spectrum. The definitions of the different categories of high column density absorber are given in Table \ref{tab:hcd_absorbers}. \textit{From top to bottom}, we show the templates for simulated results at increasing redshift [(a): \(z = 2.00\); (b): \(z = 2.44\); (c): \(z = 3.01\); (d): \(z = 3.49\); (e): \(z = 4.43\)].}
\label{fig:contaminant_ratios_templates_all_z}
\end{figure}

\begin{table*}\centering
\caption{Best-fit values of the parameters in our templates for the bias of spectra contaminated by high column density absorbers on the one-dimensional Lyman-alpha forest flux power spectrum. The template parameters are defined in Eqs.~\eqref{eq:parametric_model} and \eqref{eq:z_evolution}. Values are shown for each high column density absorber category. The definitions of the different categories of high column density absorber are given in Table \ref{tab:hcd_absorbers}.}
\label{tab:template_param_vals}
\begin{tabular}{ccccccc}
\hline
\multirow{2}{*}{Absorber category} & \multicolumn{6}{c}{Template parameter values} \\
 & \(a_0\) & \(a_1\) & \(b_0\) & \(b_1\) & \(c_0\) & \(c_1\) \\
\hline
LLS & 2.2001 & 0.0134 & 36.449 & -0.0674 & 0.9849 & -0.0631 \\
Sub-DLA & 1.5083 & 0.0994 & 81.388 & -0.2287 & 0.8667 & 0.0196 \\
Small DLA & 1.1415 & 0.0937 & 162.95 & 0.0126 & 0.6572 & 0.1169 \\
Large DLA & 0.8633 & 0.2943 & 429.58 & -0.4964 & 0.3339 & 0.4653 \\
\hline
\end{tabular}
\end{table*}

To aid incorporation in future pipelines, we have produced fits to the biases induced by contaminants in our different column density bins. The parametric form of our templates is
\begin{equation}\label{eq:parametric_model}
\frac{P_i^\mathrm{1D} (k_{||}, z)}{P_\mathrm{Forest}^\mathrm{1D} (k_{||}, z)} = \left(\frac{1 + z}{1 + z_0}\right)^{-3.55} \times \frac{1}{(a(z) e^{b(z) k_{||}} - 1)^2} + c(z),
\end{equation}
where
\begin{equation}\label{eq:z_evolution}
a(z) = a_0 \left(\frac{1 + z}{1 + z_0}\right)^{a_1};\,\,\,\,\,b(z) = b_0 \left(\frac{1 + z}{1 + z_0}\right)^{b_1};\,\,\,\,\,c(z) = c_0 \left(\frac{1 + z}{1 + z_0}\right)^{c_1};
\end{equation}
and the pivot redshift \(z_0 = 2.00\). \([a_0, a_1, b_0, b_1, c_0, c_1]\) are free parameters that we fit simultaneously in \(k_{||}\) and \(z\) space for each absorber category \(i \in\) \{LLS, sub-DLA, small DLA, large DLA\}.  We fit using the Levenberg-Marquardt algorithm \citep{citeulike:10796881,citeulike:928708}\footnote{We were able to further validate our modelling by initially fitting using a subset of the available redshift slices and using this preliminary fit to predict the results at \(z = 3.01\). We found the model to accurately predict the results at this intermediate redshift, acting as a form of successful blind test for our model. Our final best-fit parameters use all available data.}.

Figure \ref{fig:contaminant_ratios_templates_all_z} shows the result of these fits (dashed lines) with the raw ratios measured from the simulation (solid lines); the corresponding parameter values are given in Table \ref{tab:template_param_vals}. These can be used to reconstruct a final model for the bias of spectra containing high column density absorbers by using Eq.~\eqref{eq:multiplicative_bias}.
The model described by Eq.~\eqref{eq:parametric_model} characterises the results we have measured in our simulations and through Eq.~\eqref{eq:z_evolution} allows interpolation of our results to intermediate redshifts that we have not explicitly probed. (Use of the model outside the limits of redshift and scale we have considered would constitute an extrapolation, but this should not be necessary since our measurements bracket the main redshifts and scales of interest to Lyman-alpha forest studies.) No strong physical meaning should be attached to its terms, although we can motivate the first term on the right-hand side of Eq.~\eqref{eq:parametric_model} as being the (reciprocal of the) main term of the redshift evolution of \(P_\mathrm{Forest}^\mathrm{1D} (k_{||}, z)\) as found by \citet{2013A&A...559A..85P} (using a maximum likelihood estimator). In this way, the parametric form isolates the redshift evolution from \(P_\mathrm{Forest}^\mathrm{1D} (k_{||}, z)\) and then fits the residual redshift evolution using the terms in Eq.~\eqref{eq:z_evolution}. The best-fit values of the exponents in Eq.~\eqref{eq:z_evolution} (as given in Table \ref{tab:template_param_vals}) are small, indicating that most of the redshift evolution can indeed be ascribed to the expected cosmological evolution of \(P_\mathrm{Forest}^\mathrm{1D} (k_{||}, z)\).

Our results are dependent on the length of our simulated spectra. This manifests in the value of the constant that the ratios \(P_i^\mathrm{1D} (k_{||}, z) / P_\mathrm{Forest}^\mathrm{1D} (k_{||}, z)\) have at high \(k_{||}\), which is set by the fraction of the length of contaminated spectra which are unaffected by damping wings and contain only Lyman-alpha forest. Since the incidence rates of high column density absorbers are such that one per contaminated spectrum is most likely, a longer spectrum will have a larger fraction that is uncontaminated, causing the constant value at high \(k_{||}\) to rise with spectrum length. However, in an analysis of observational data this will be absorbed into a free parameter. We have used a parametric form for our templates such that all this dependency is measured by the term \(c(z)\)\footnote{It can then be understood why we do not factor out the redshift evolution of  \(P_\mathrm{Forest}^\mathrm{1D} (k_{||}, z)\), as we do for the first term on the right-hand side of Eq.~\eqref{eq:parametric_model}.}. By inserting Eq.~\eqref{eq:parametric_model} into Eq.~\eqref{eq:multiplicative_bias}, it can be seen that the term \(c(z)\) is degenerate with \(\alpha_\mathrm{Forest} (z)\) and hence these terms can be combined and allowed to vary. It follows that the full parametric form of our model for the effect of high column density absorbers on the 1D Lyman-alpha forest flux power spectrum is
\begin{equation}\label{eq:full_parametric_model}
\begin{split}
P_\mathrm{Total}^\mathrm{1D} (k_{||}, z) = P_\mathrm{Forest}^\mathrm{1D} &(k_{||}, z)\,\left[\alpha_0 (z)\vphantom{\frac{0}{0}}\right.\\
&\left. + \sum_{i \neq \mathrm{Forest}} \alpha_i (z) \left(\frac{1 + z}{1 + z_0}\right)^{-3.55} \times \frac{1}{(a(z) e^{b(z) k_{||}} - 1)^2}\right].
\end{split}
\end{equation}
When using this model in inference from the 1D Lyman-alpha forest power spectrum \(P_\mathrm{Forest}^\mathrm{1D} (k_{||}, z)\), it will be necessary to vary five free parameters \(\alpha_0\) and \(\alpha_i\), where \(i\) indexes each high column density absorber category. In this way, the column density, scale and redshift dependence of the effect of high column density absorbers is fully determined by our templates, while the relative impact of each absorber category is fitted since this is specific to the survey at hand, as well as the details of any clipping of damping wings that changes the survey CDDF. (See \S~\ref{sec:discussion} for more discussion of these details.) Note that the parameter \(\alpha_0\) is degenerate with factors that rescale the mean flux and could be omitted in an end-to-end analysis.

\begin{figure}
\includegraphics[width=\columnwidth]{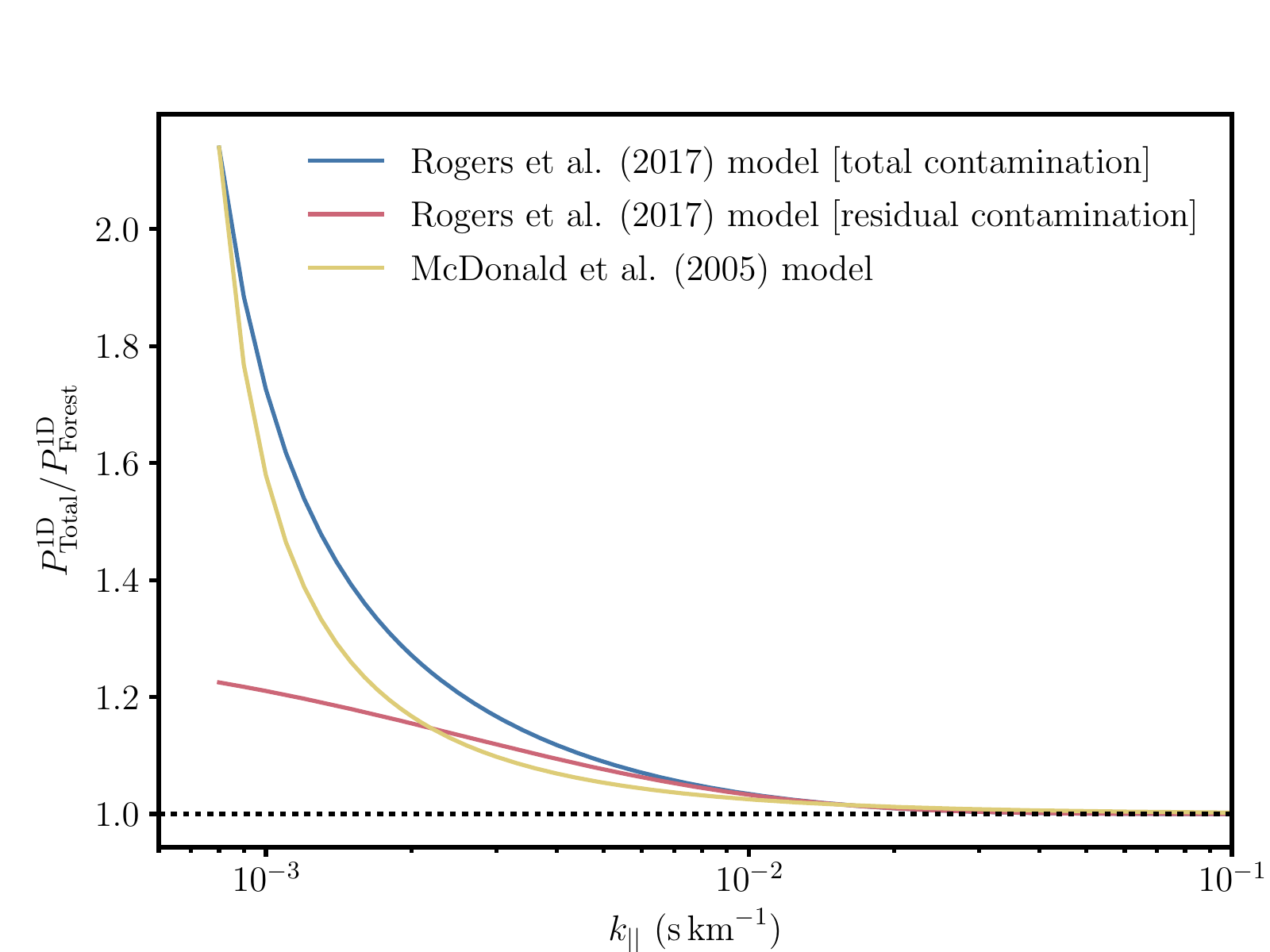}
\caption{A comparison of the existing multiplicative bias model \citep{2005MNRAS.360.1471M,2015JCAP...11..011P} for the effect of high column density absorbers on the 1D Lyman-alpha forest power spectrum and the model constructed in this paper using our results from hydrodynamical simulations. For our model, we show an example weighting of the different absorber categories for the full contamination from high column density absorbers on our simulated ensemble of spectra; and an example based on a possible ``residual'' contamination after the clipping of DLAs (\ie only LLS and sub-DLAs remaining). For comparison, the model of \citet{2005MNRAS.360.1471M} is rescaled to have the same amplitude on the largest and smallest scales considered.}
\label{fig:model_comparison}
\end{figure}

Figure \ref{fig:model_comparison} compares the model we have constructed to the existing model presented in \citet{2015JCAP...11..011P} and based on the results in \citet{2005MNRAS.360.1471M}. There is broad agreement between the existing model and our model for the total contamination of high column density absorbers, although our model is less steep in its scale dependence. We also show our model applied to a possible ``residual'' contamination, \ie under the assumption that all DLAs are identified and clipped out in an analysis, leaving only contamination from LLS and sub-DLAs (\eg as assumed by \citealt{2017arXiv170200176B}). The model for this lower column density residual contamination has a shallow scale dependence that the model of \citet{2005MNRAS.360.1471M} is unable to characterise. The use of our more flexible model will avoid potential biases due to mischaracterisation of the scale dependence of the residual contamination, thus improving estimation of cosmological effects such as massive neutrinos or the tilt of the primordial power spectrum.

We now discuss the prior probability distributions that can be adopted for \(\alpha_i (z)\) in any inference using the model we have presented. The \(\alpha_i (z)\) are technically not independent parameters, but are each related to integrals of the {\sc Hi} CDDF for a particular survey over the appropriate column density ranges (and absorption distance per sightline). The effect of spectrum clipping which changes the survey CDDF can be modelled by applying a weighting function to the CDDF, which down-weights higher column densities, which are easier to spot and remove. If one wanted to reduce the dimensionality of these nuisance parameters, in particular in redshift space, they could be replaced by a parameterisation which quantifies deviations from the expected redshift evolution of the CDDF with only one or two parameters (rather than a parameter for each redshift bin considered). We leave the details of the construction of prior distributions to individual analyses, since the precise considerations will be survey-specific.

To conclude this section, we present a summary of the steps required to incorporate our final model for the effect of high column denisty absorbers into future 1D Lyman-alpha forest analyses:
\begin{itemize}
\item Our model describes the effect of quasar spectra contaminated by high column density absorbers as a multiplicative bias to the 1D Lyman-alpha forest flux power spectrum, as given by Eq.~\eqref{eq:full_parametric_model}. It can therefore be incorporated into a pipeline at the stage of flux power spectrum interpretation to marginalise over effects of these absorbers.  
\item The free parameters are \(\alpha_i (z)\), where \(i\) indexes different categories of high column density absorber (as given in Table \ref{tab:hcd_absorbers}). Our model is of use to any Lyman-alpha forest survey that contains spectra which may be contaminated by high column density absorbers (both LLS and DLAs). The relative impacts of different categories of high column density absorbers will be determined in the estimation of posterior distributions of these nuisance parameters. While normalisation is necessarily floating, the model fully specifies the scale, column density and redshift dependence of the effect of high column density absorbers, using the results we have measured from hydrodynamical simulations.
\item In a survey which does not clip its quasar spectra, strong priors can be given for the free parameters of our model, based on the expected or measured {\sc Hi} CDDF.
\item In a survey which does clip its quasar spectra in an attempt to remove high column density absorbers (and therefore changes the survey CDDF), strong priors can still be given for our model parameters, assuming a model can be constructed for the effect of the clipping process on the CDDF. This will constitute some re-weighting of the CDDF.
\item In order to reduce the dimensionality of our nuisance parameters, rather than having a separate parameter for each redshift bin in a given analysis, one could parameterise the redshift evolution by a simple deviation from the CDDF with only one or two numbers.
\end{itemize}

\section{Conclusions}
\label{sec:concs}

We have used a cosmological hydrodynamical simulation \citep[Illustris;][]{2014Natur.509..177V,2015A&C....13...12N} to investigate the effect of high column density absorbing systems of neutral hydrogen and their associated damping wings on the 1D Lyman-alpha forest flux power spectrum. We find that the effect of high column density absorbers on the Lyman-alpha forest flux power spectrum is a strong function of column density. Accounting for this change in scale-dependence with column density will remove a source of bias in cosmological inference from the Lyman-alpha forest.  Previous models \citep{2015JCAP...11..011P} combine the effect of all high column density absorbers together (\ie all neutral hydrogen column densities \(N(\textsc{Hi}) > 1.6 \times 10^{17}\,\mathrm{atoms}\,\mathrm{cm}^{-2}\)) based on the column density distribution function (CDDF) in the raw spectra \citep{2006ApJS..163...80M}. However, the damping wings of some high column density absorbers are clipped out in the final analysis \citep{2013AJ....145...69L}, which preferentially removes higher density systems (because they are easier to spot) and changes the column density distribution in the residual contamination.  Our results apply for both clipped and unclipped survey spectra, since we separately model the effect for different column densities of the dominant absorber, allowing us to accurately account for the contamination in the 1D flux power spectrum. We discuss in \S~\ref{sec:templates} the practicalities of employing our model in future analyses.

The shape and amplitude of the distortions in the power spectrum due to a damped absorber depend on its column density because they are driven by the width of the damping wings; \ie the dominant effect is a ``one-halo'' term. We defer investigation of potential ``two-halo'' terms to future work where we measure the effect of high column density absorbers on the 3D Lyman-alpha forest flux power spectrum.

We anticipate that our model will help realise forecasted cosmological constraints from upcoming surveys like DESI. \Eg \citet{2014JCAP...05..023F} forecast that DESI will have the constraining power to make a \(\sim\) three-sigma detection of the sum of neutrino masses (in combination with \textit{Planck} CMB data); and they show the power of the 1D Lyman-alpha forest power spectrum in probing the primordial power spectrum, \eg halving the one-sigma error on the running of the spectral index, with implications for inflationary models. It will be necessary to use the models we have presented here, alongside carefully constructed priors on the residual CDDF, to remove degeneracies between the effect of high column density absorbers and cosmological effects.

\section*{Acknowledgements}
\label{sec:ack}

KKR, SB, HVP and BL thank the organisers of the COSMO21 symposium in 2016, where this project was conceived. KKR was supported by the Science and Technology Facilities Council (STFC). SB was supported by NASA through Einstein Postdoctoral Fellowship Award Number PF5-160133. HVP was partially supported by the European Research Council (ERC) under the European Community's Seventh Framework Programme (FP7/2007-2013)/ERC grant agreement number 306478-CosmicDawn. AP was supported by the Royal Society. AFR was supported by an STFC Ernest Rutherford Fellowship, grant reference ST/N003853/1. BL was supported by NASA through Einstein Postdoctoral Fellowship Award Number PF6-170154.

\bibliographystyle{mymnras_eprint}
\bibliography{dla_bias_1D}

\appendix
\section{One-dimensional flux power spectrum of a Voigt profile}
\label{sec:voigt}

\begin{figure}
\includegraphics[width=\columnwidth]{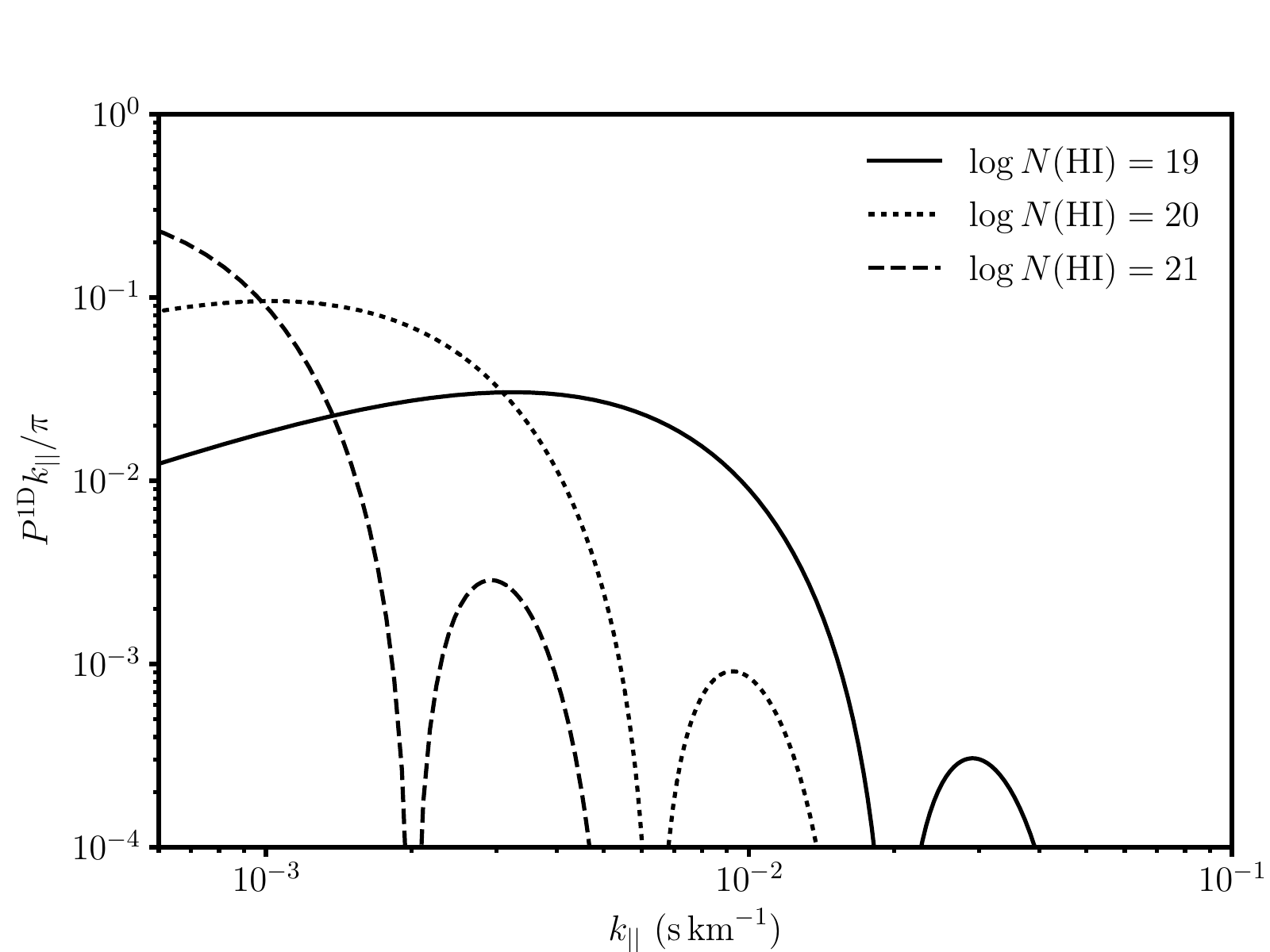}
\caption{The one-dimensional flux power spectra of Voigt profiles of broadened Lyman-alpha absorption lines as generated by different column densities of neutral hydrogen \(N(\textsc{Hi})\), as a function of line-of-sight scale \(k_{||}\) (the units of \(N(\textsc{Hi})\) are \(\mathrm{atoms}\,\mathrm{cm}^{-2}\)).}
\label{fig:voigt_power}
\end{figure}

As discussed in \S~\ref{sec:hcd_absorbers}, the broadened absorption lines of high column density absorbers are usually modelled by a Voigt profile. A Voigt profile is a convolution of a Lorentzian profile and a Gaussian profile. It therefore appropriately models the combination of the main physical processes that broaden atomic transition lines: the Lorentzian profile from \eg natural or collisional broadening and the Gaussian profile from \eg Doppler broadening. The optical depth as a function of wavelength \(\tau (\lambda)\) is the product of the line-of-sight column density \(N\) and the atomic absorption coefficient \(\alpha (\lambda)\) \citep[\eg][]{HUMLICEK1979309}\footnote{Eq.~\eqref{eq:voigt_tau} is valid in SI units.}:
\begin{equation}\label{eq:voigt_tau}
\tau (\lambda) = N \alpha (\lambda) = N \frac{\sqrt{\pi} e^2}{4 \pi \epsilon_0 m_\mathrm{e} c^2} \frac{f \lambda_\mathrm{t}^2}{\Delta\lambda_\mathrm{D}} u(x,y),
\end{equation}
where the fundamental physical constants have their usual meaning, \(f\) is the oscillator strength of the atomic transition, \(\lambda_\mathrm{t}\) is the transition wavelength and the Doppler wavelength ``shift'' associated with a gas of temperature \(T\) for an ion of mass \(m_\mathrm{ion}\),
\begin{equation}\label{eq:doppler_shift}
\Delta\lambda_\mathrm{D} = \frac{\lambda_\mathrm{t}}{c} \left(\frac{2 k_\mathrm{B} T}{m_\mathrm{ion}}\right)^\frac{1}{2}.
\end{equation}
\(u(x,y)\) is an unnormalised form of the Voigt function (the normalisation is already expressed in the pre-factors of Eq.~\eqref{eq:voigt_tau}), specifically the real part of the Faddeeva function:
\begin{equation}\label{eq:faddeeva_func}
w(z) = e^{-z^2} \mathrm{erfc}(-iz) = u(x,y) + iv(x,y),
\end{equation}
where \(\mathrm{erfc}(x)\) is the complementary error function and \(z = x + iy\). \(x\) and \(y\) are respectively the wavelength difference from the line centre \(\lambda_\mathrm{c}\) and the natural width of the transition, in units of the Doppler shift:
\begin{equation}\label{eq:x_y}
x(\lambda) = \frac{\lambda - \lambda_\mathrm{c}}{\Delta\lambda_\mathrm{D}};\,\,\,\,\,y = \frac{\Gamma \lambda_\mathrm{t}^2}{4 \pi c} \frac{1}{\Delta\lambda_\mathrm{D}},
\end{equation}
where \(\Gamma\) is the damping constant of the transition, \ie the inverse of the time scale for the electron to remain in the upper level of the transition in the vacuum. For the Lyman-alpha transition, \(f = 0.4164\), \(\lambda_\mathrm{t} = 1215.67 \angstrom\), \(m_\mathrm{ion} = m_\mathrm{proton}\) and \(\Gamma = 6.265 \times 10^8\,\mathrm{Hz}\) \citep{0067-0049-151-2-403}. For the column densities that we consider, we assume a gas temperature \(T \approx 10^4 K\). In order to calculate the 1D flux power spectrum arising from these Voigt profiles, the same procedure is followed as in \S~\ref{sec:pow_spectrum}, \ie we form flux spectra and carry out a Fourier transform. We transform from wavelengths to velocities by \(\Delta v / c = \Delta \lambda / \lambda\).

Figure \ref{fig:voigt_power} shows the 1D flux power spectra of Voigt profiles as given by Eq.~\eqref{eq:voigt_tau} for the Lyman-alpha absorption line for three different column densities of neutral hydrogen \(N(\textsc{Hi}) = [10^{19}, 10^{20}, 10^{21}]\,\mathrm{atoms}\,\mathrm{cm}^{-2}\), spanning the column densities for LLS and DLAs. This figure should be compared with Fig.~\ref{fig:contaminant_power} in \S~\ref{sec:results}, which shows the 1D flux power spectra we have measured in the hydrodynamical simulations. The trends in Fig.~\ref{fig:voigt_power} broadly support the arguments made in \S~\ref{sec:discussion}, relating the large-scale power spectrum of simulated spectra contaminated by high column density absorbers to the power spectrum of relevant Voigt profiles. The shape of the large-scale power spectrum of the Voigt profiles is similar in amplitude and scale-dependence as the excesses on large scales for the 1D flux power spectra of simulated spectra in high column density absorber categories. Moreover, these excesses get steeper, increase in amplitude and become prominent on larger scales for higher column densities, both in the simulated and analytic spectra. This reflects the fact that a higher column density means wider damping wings and so correlations on larger scales. In the analytic power spectra in Fig.~\ref{fig:voigt_power}, we observe oscillations in the power spectrum on smaller scales that rapidly decrease in amplitude. These are not observed in the fully-simulated power spectra since the oscillations are orders of magnitude lower in amplitude than the flux power spectrum of residual Lyman-alpha forest (see Fig.~\ref{fig:contaminant_power}). Furthermore, in our results, we are effectively averaging over a number of column densities in each column density bin (or absorber category) that we consider; this will have the additional effect of averaging out these smaller-scale oscillations in the power spectrum to form a smoother scaling.

\end{document}